\documentclass[aps,prb,twocolumn,groupedaddress,showpacs]{revtex4}
\usepackage{epsfig}
\usepackage{color}
\newcommand{\beq}{\begin{equation}}
\newcommand{\eeq}{\end{equation}}
\newcommand{\bea}{\begin{eqnarray}}
\newcommand{\eea}{\end{eqnarray}}

\newcommand{\bary}{\begin{array}}
\newcommand{\eary}{\end{array}}
\newcommand{\benum}{\begin{enumerate}}
\newcommand{\eenum}{\end{enumerate}}
\newcommand{\bitem}{\begin{itemize}}
\newcommand{\eitem}{\end{itemize}}
\begin{document}

\title{Canonical Trajectories and Critical Coupling of the
  Bose-Hubbard Hamiltonian in a Harmonic Trap}

\author{G.G. Batrouni$^1$, H. R. Krishnamurthy$^{2,3}$,
  K. W. Mahmud$^2$, V.G. Rousseau$^4$, and R.T. Scalettar$^2$}

\affiliation{$^1$INLN, Universit\'e de Nice-Sophia Antipolis, CNRS;
  1361 route des Lucioles, 06560 Valbonne, France}

\affiliation{$^2$Physics Department, University of California, Davis,
  California 95616, USA}

\affiliation{$^3$Centre for Condensed Matter Theory, Department of
  Physics, Indian Institute of Science, Bangalore 560012, India }

\affiliation{$^4$Lorentz Institute, Leiden University, P. O. Box 9506, 2300
RA  Leiden, The Netherlands}

\begin{abstract}
Quantum Monte Carlo (QMC) simulations and the Local Density
Approximation (LDA) are used to map the constant particle number
(canonical) trajectories of the Bose Hubbard Hamiltonian confined in a
harmonic trap onto the $(\mu/U,t/U)$ phase diagram of the uniform
system.  Generically, these curves do not intercept the tips of the
Mott insulator (MI) lobes of the uniform system. This observation
necessitates a clarification of the appropriate comparison between
critical couplings obtained in experiments on trapped systems with
those obtained in QMC simulations.  The density profiles and
visibility are also obtained along these trajectories. Density
profiles from QMC in the confined case are compared with LDA results.
\end{abstract}

\maketitle

The bosonic Hubbard model was first introduced \cite{fisher89} in the
context of disordered superconductors where the superfluidity of
preformed Cooper pairs competes with Mott insulator and Bose glass
phases.  Considerable numerical work followed the original analytic
treatment.  When there is no disorder, Quantum Monte Carlo (QMC)
studies \cite{batrouni90,prokofev98} obtained quantitative values for
the critical coupling of the superfluid-Mott insulator (SF-MI)
transition at commensurate filling in one dimension, which were in
good quantitative agreement with series expansion \cite{freericks96}
and density matrix renormalization group calculations
\cite{kuhner98,kuhner00}.  The critical point is now known in $d=2$ to
a very high accuracy \cite{prokofev07}.

Over the last decade, it became clear that trapped ultra-cold atoms
provide an alternate, and more controllable, experimental realization
of the bosonic Hubbard model \cite{jaksch98}.  Indeed, the possibility
of a quantitative comparison of theoretical and experimental values
for the critical point has been suggested.  A recent experimental
paper \cite{spielman07} has offered the first such benchmark in $d=2$.

However, a significant obstacle exists for such a direct comparison:
The confining potential produces spatial inhomogeneities and a
coexistence of SF and MI phases \cite{batrouni02}. This naturally
leads to the question as to what ``critical coupling" is being
accessed in the experiments.  Is it the coupling at which ``Mott
shoulders'' begin to develop about a SF core?  Or is it the coupling
at which a Mott region pervades the entire central region of the trap?
In this paper, we provide a detailed quantitative analysis of this
issue.  Specifically, using the Local Density Approximation (LDA) and
QMC simulations, we study, for fixed particle numbers, the evolution
of the density profiles of the trapped system as a function of the
interaction strength and map those ``canonical trajectories'' onto the
phase diagram of the uniform system.  We also show data for the
visibility \cite{bloch05,sengupta05}. These measurements allow us to
connect the critical points obtained in QMC with those that can be
seen in experiment.

The QMC results presented here were obtained using two different
algorithms. In the first \cite{hirsch82}, the imaginary time $\beta$
is discretized leading to a path integral for the partition function
on a rigid space-imaginary time grid with local world line updates. In
the second \cite{beard96,prokofev96,rombouts06}, imaginary time is
continuous and there are no Trotter errors associated with
discretization.  Bosonic world-line updates can be non-local, and, as
a consequence, the Green's function can be measured at all
separations. The two algorithms give consistent results for all
physical quantities calculated such as the density profiles and
superfluid density.

The one dimensional bosonic Hubbard Hamiltonian is,
\begin{eqnarray}
\nonumber
H &=&-t\sum_{i} \left(a^{\dagger}_i a_{i+1} + a^{\dagger}_{i+1}
a_i\right) - \mu\sum_i n_i\\ 
& & + V_T\sum_i x_i^2\ n_i + \frac{U}{2}\sum_i n_i(n_i-1) \,\,.
\label{hubham}
\end{eqnarray}
Here $i=1,2,\cdots, L$ where $L$ is the number of sites and
$x_i=a[i-L/2]$ is the coordinate of the $i$th site as measured from
the center of the system. We choose the lattice constant $a=1$.  The
hopping parameter, $t$, sets the energy scale; in what follows we set
$t=1$, i.e., all energies are measured in units of $t$.
$n_i=a^\dagger_i a_i$ is the number operator, and
$[a_i,a^\dagger_j]=\delta_{ij}$ are bosonic creation and destruction
operators. $V_T$ is the curvature of the trap, and the repulsive
contact interaction is given by $U$. The chemical potential, $\mu$,
controls the average number of particles.

The bosonic-Hubbard Hamiltonian can also be simulated in the canonical
ensemble at fixed particle number $N_b$. Indeed this is essential in
order to make contact with experiments. In the homogeneous case,
$V_T=0$, the phase diagram is a function of the density, $N_b / L^d$,
and the interaction $U/t$ where $d$ is the dimensionality of the
system.  It was emphasized recently \cite{rigol04} that a similar
lattice size independent formulation can be made in the confined case
by using a rescaled length $\xi_i\equiv x_i/\xi$ with $\xi =
\sqrt{t/V_T}$. Then, density profiles and the resulting phase diagram
depend on $N_b$ and $V_T$ only via the combination $\tilde \rho =
N_b/\xi^d$, called the ``characteristic density".

\begin{figure}[h]
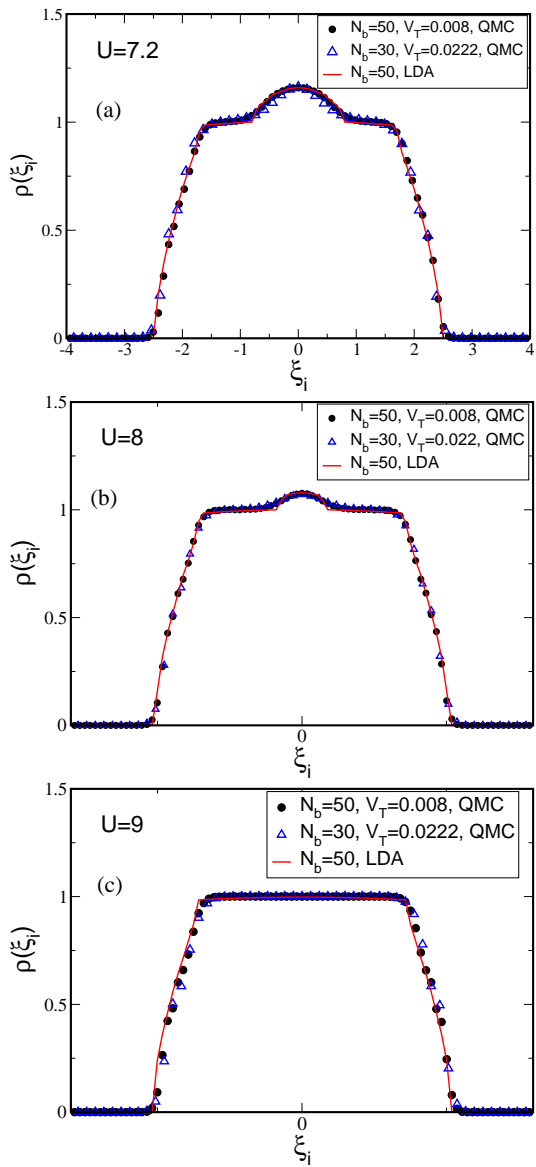

\centerline{\epsfig{figure=fig1a.eps,
width=7cm,clip}}
\centerline{\epsfig{figure=fig1b.eps,
width=7cm,clip}}
\centerline{\epsfig{figure=fig1c.eps,
width=7cm,clip}}
\caption{(Color online) Density profiles {\it vs} the rescaled
  position, $\xi_i$, in $1d$.  The solid lines are obtained using QMC
  for the uniform system combined with the LDA to include the trap.
  The symbols are the results of QMC done directly on the confined
  system.  The characteristic density $\tilde \rho= 4.47$ and $U=7.2,
  8.0, 9.0$.  We also show, in the three panels, profiles
  for two different particle numbers, $N_b=50,30$ but with the same
  $\tilde \rho= 4.47$.}
\label{densityprofile50}
\end{figure}

One simple way to understand the role of the characteristic density
and to infer the properties of the trapped system is the Local Density
Approximation (LDA) \cite{bergkvist04} in which the density at a
particular location $x_i$ in the trapped system is assumed to be given
by the density of a uniform system with chemical potential equal to
the ``local chemical potential'' $\mu_i \equiv \mu - V_T x_i^2$ at
that location.  In other words, for a trapped $1d$ system,
\begin{equation}
\rho(x_i) \equiv \langle n_i \rangle_{V_T} = \rho_{1d}^0(\mu_i; U),
\label{ldaeq}
\end{equation}
where $\rho_{1d}^0(\mu ; U) \equiv \langle n_i \rangle_{V_T = 0}$ is
the density for the $1d$ bosonic Hubbard model in the homogeneous
case. For a given desired $N_b$, the requisite
chemical potential $\mu$ in the presence of the trap, which is also
the {\it local} chemical potential at the center of the trap, {\em
is determined by the condition},
\begin{equation}
N_b = \sum_i \langle n_i \rangle_{V_T} = \sum_i \rho_{1d}^0(\mu -
V_T x_i^2 ; U), \label{mu-eqn-1}
\end{equation}
and is therefore implicitly a function of $N_b, U$ and $V_T$. In
principle, one can use $\rho_{1d}^0(\mu ; U)$ as determined by QMC in
the uniform case, together with Eq.~(\ref{mu-eqn-1}) to determine
$\mu$. Within the LDA the density profile $\rho(x_i)$ is then
completely determined, and can be compared with results obtained
directly from simulations with a trap potential to determine the
accuracy of the LDA, as discussed below.  Furthermore,
Eq.~(\ref{mu-eqn-1}) provides a useful guide to understanding the
trajectories in the $(\mu, U)$ plane that are traversed in
experimental investigations such as in Ref.~\onlinecite{spielman07},
since they are typically done at fixed $N_b$ and varying $t$ by
varying the depth of the optical potential (which, however, also
changes the trap potential).

\begin{figure}[h]
\centerline{\epsfig{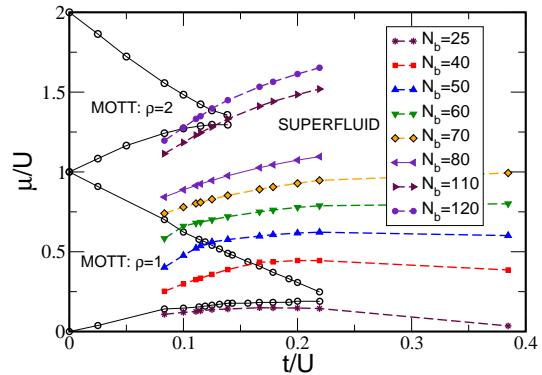}}
\caption{(Color online) Canonical (constant particle number) flows in
the $(\mu/U,t/U)$ plane at fixed $V_T=0.008$.  Characteristic densities
vary from $\tilde \rho = 2.24$ for $N_b=25$ (lowest curve) to $\tilde
\rho = 10.73$ for $N_b=120$ (highest curve).  }
\label{phasediag}
\end{figure}

Better insight into the nature of such {\em canonical (constant $N_b$)
trajectories} is obtained by approximating the sum in
Eq.~(\ref{mu-eqn-1}) as an integral. This should be a reasonable
approximation when the local chemical potential changes slowly from
site to site, i.e, in the same regime where LDA is expected to be
valid. In the one dimensional case\cite{higher-dim} one has, \beq N_b
= 2 \int_0^{\infty} dx \, \rho_{1d}^0(\mu - V_T x^2 ; U).  \eeq
Changing the integration variable to $\mu_x \equiv \mu - V_T x^2$,
this equation can be rewritten as\cite{higher-dim},
\begin{equation}
N_b \sqrt{V_T} \equiv \tilde{\rho} = \int_{-\infty}^{\mu}
{\frac {d\mu_x \, \rho_{1d}^0( \mu_x; U)} {\sqrt{\mu-\mu_x}}} \equiv
I_1 (\mu; U).
\label{ch-den-eqn}
\end{equation}
$I_1$ is entirely determined from the solution of the homogeneous
problem. The chemical potential in the presence of the trap, $\mu$, is
determined by inverting Eq.~(\ref{ch-den-eqn}). Note the natural
appearance of the characteristic density ${\tilde \rho}$ on the left
side of Eq.~(\ref{ch-den-eqn}).  Clearly, $\mu$, and hence the density
profile expressed as a function of $x/\xi$, depend only on ${\tilde
\rho}$ and not on $N_b$ and $V_T$ separately. Needless to say, $I_1$
can also be computed directly by evaluating the sum in
Eq.~(\ref{mu-eqn-1}) using the simulation results for $\rho_{1d}^0$.

Thermodynamic stability implies that $\rho_{1d}^0$, and hence
$I_1(\mu;U)$, are monotonically increasing functions\cite{rhomu} of
$\mu$.  For $\tilde{\rho} < I_1(\mu_1^-(U);U) \equiv
\tilde{\rho}_1^{\,-}(U)$, $\mu$, and hence $\mu_i$, are less than
$\mu_1^-(U)$, the chemical potential at which the first Mott lobe is
reached from below.  Therefore $\rho(x_i) < 1$, and all sites are
sampling the SF region in the phase diagram below the first Mott lobe
(if $U > U_c$).

The density profile is very different when $U$ , $V_T$ and $N_b$ are
such that $\tilde{\rho}$ is larger than $\tilde{\rho}_1^{\,-}(U)$.
Then $\mu > \mu_1^-(U)$ and therefore a flat Mott plateau with
$\rho(x_i) = 1$ appears in the central region of the system, extending
over sites $i$ for which $\mu_i \ge \mu_1^-(U)$. For sites outside
this plateau, $\rho(x_i)<1$ and the system is locally in the SF phase.

If the trap potential is increased so as to squeeze the particles
towards the center of the cell (or if $N_b$ is increased),
$\tilde{\rho}$ and $\mu$ increase. For $\tilde{\rho} >
I_1(\mu_1^+(U);U) \equiv \tilde{\rho}_1^{\,+}(U)$, one has $\mu >
\mu_1^+(U)$, the chemical potential at which the first Mott lobe is
reached from above. In this case the central sites of the system are
in the superfluid region above the Mott lobe, with $\rho(x_i) > 1$,
surrounded by MI shoulders where $\rho(x_i)=1$, in turn surrounded by
SF regions as the edges of the system are reached
(Fig.~\ref{densityprofile50}(a)).

In the regime $\tilde{\rho}_1^{\,-}(U) < \tilde{\rho} <
\tilde{\rho}_1^{\,+}(U)$, as is easily verified from
Eq.~(\ref{ch-den-eqn}), $\mu$ is determined by the equation,
\begin{equation}
\mu - \mu_1^-(U) = [\tilde{\rho} - \tilde{\rho}_1^{\,-}(U)]^2 / 4.
\label{mu-eqn-2}
\end{equation}
Hence the two threshold values of $\tilde{\rho}$ in the presence of
the trap and the threshold chemical potentials for the Mott transition
in the homogeneous case are related via,
\begin{equation}
\mu_1^+(U) - \mu_1^-(U) = [\tilde{\rho}_1^{\,+}(U) -
  \tilde{\rho}_1^{\,-}(U)]^2 / 4.
\label{ch-den-mu-relation}
\end{equation}
For larger values of $\tilde{\rho}$ in large systems with a small
$V_T$, one can access transitions involving the higher Mott lobes
\cite{batrouni02}.

In Fig.~\ref{densityprofile50} we compare the density profiles
obtained from direct QMC simulations of the trapped system with those
inferred from the LDA and QMC simulations of the uniform system.  The
LDA generally provides an accurate description of the density profiles
except at those locations in the trap where a changeover from
superfluid to Mott insulator region is occurring. This is clear in
Fig.~\ref{densityprofile50} where as one goes from SF to MI regions,
the transition is much sharper for the LDA curves. This, of course, is
a vestige of the true quantum phase transition present in the
unconfined system on which the LDA method is based.
Figure~\ref{densityprofile50} shows profiles for two different pairs
of $(N_b,V_T)$ which have the same characteristic density.  They are
seen to coincide almost perfectly, validating the use of $\xi_i$ and
$\tilde \rho$ to describe the physics in a scale-independent way.

\begin{figure}[h]
\centerline{\epsfig{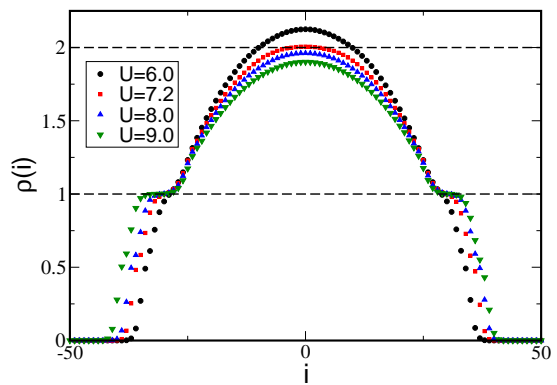}}
\caption{(Color online) Density profiles along the $N_b=110$ ($\tilde
  \rho = 9.84$) trajectory.  This value just clips the tip of the
  uniform system $\rho=2$ lobe, as seen in Fig.~\ref{phasediag}. The
  dashed lines are to draw attention to the $\rho=1,2$ values where
  the MI develops.}
\label{densityprofile110}
\end{figure}

Figure~\ref{phasediag} shows the canonical trajectories corresponding
to $\mu(\tilde{\rho},U)$ for fixed $\tilde{\rho}$, obtained from
Eq.~(\ref{mu-eqn-1}), superimposed on the phase diagram of the uniform
system. Each trajectory is at constant $N_b$ and, therefore, constant
${\tilde \rho}$ when $V_T$ is fixed, and shows where the trapped
system sits in the phase diagram of the uniform system when the LDA is
used in combination with QMC. For example, for the confined system
values $N_b=50$, $V_T = 0.008$ and $U=9.0$, $\mu(\tilde{\rho},U)$,
lies well within the $\rho=1$ Mott lobe and the system should be a
Mott insulator according to this mapping.
Figure~\ref{densityprofile50}(c) shows the true density profile
obtained with QMC directly with a trap and we see that, indeed, the
confined system is a MI, except for the edges which always have
$\rho(x_i)<1$. On the other hand, staying on the same trajectory,
$N_b=50$, but with $U=7.2$, the $\mu(\tilde{\rho},U)$ lies in the SF
phase above the Mott lobe leading us to predict the central region of
the trapped system to be SF with $\rho(x_i) > 1$, as indeed confirmed
by Fig.~\ref{densityprofile50}(a).  A second example of this
evolution, for $N_b=110$, which just clips the top of the $\rho=2$
Mott lobe, is given in Fig.~\ref{densityprofile110}, with similar
conclusions. Notice that as $U$ increases, if a trajectory enters,
say, the $\rho=2$ Mott lobe, it will leave it eventually upon further
increases in $U$.  Such a trajectory will eventually enter the
$\rho=1$ Mott lobe which it can never leave.

It is important to note that different trajectories intersect the Mott
lobes at different $(\mu/U,t/U)$ points and in general {\it not} at
the tip. Thus, Fig.~\ref{phasediag} emphasizes the central point of
this paper, namely that {\it both } the particle number and confining
potential need to be considered together in determining the `critical
point' of the trapped boson Hubbard Hamiltonian.  In particular, in
order to access $U_c$ in an experiment, the characteristic density
also has to be tuned to its appropriate critical value.  In the case
of a 1-d trapped system we are considering in this paper,
${\tilde{\rho}}_c \simeq 2.7$.

\begin{figure}[h]
\centerline{\epsfig{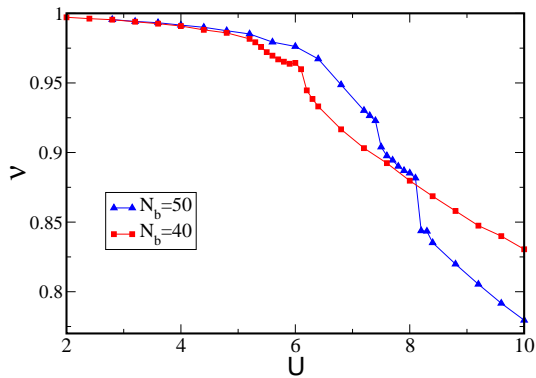}}
\caption{(Color online) The visibility along the $N_b=40$ and $N_b=50$
trajectories (corresponding to $\tilde \rho= 3.58$ and $4.47$ with
$V_T=0.008$).  For $N_b=50$ the kink at $U=7.5$ is associated with the
presence of well-formed Mott shoulders.  The second kink at $U=8.2$
corresponds to the formation of a full Mott phase throughout the
center of the trap ($V_T=0.008$).  }
\label{visib}
\end{figure}

Our understanding of the relation between the density profiles and the
``flow diagram'' of canonical trajectories is made complete by
examining the visibility ${\cal V}$, which is known to be a sensitive
measure of the behavior of the density profiles
\cite{bloch05,sengupta05}.  For $N_b=50$ ($\tilde \rho= 4.47$), ${\cal
V}$ has two kinks at $U=7.5$ and $U=8.2$ which indicate respectively
the appearance of well-formed Mott shoulders surrounding a SF interior
and then the total disappearance of superfluidity at the trap center
and the establishment of MI throughout (Fig.~\ref{visib}).  It is seen
from Fig.~\ref{phasediag} that the second, larger, of these two values
corresponds very well to the coupling where the $N_b=50$ trajectory
enters the uniform system Mott lobe.

In summary, in this paper we have shown that for fixed particle
number, the ``critical coupling'' associated with destruction of
superfluidity and onset of Mott behavior depends on the characteristic
density $\tilde \rho$.  In fact, this observation is also implicit in
the ``state diagram'' of [\onlinecite{batrouni02}] in which the
boundaries between phases at fixed $V_T$ were shown to depend on
$N_b$.  Using the local density approximation we explicitly
constructed the trajectories in the ($\mu/U$, $t/U$) plane which
correspond to constant $\tilde \rho$, and quantified their points of
entry into the Mott lobe of the uniform system.  This construction
should allow experimentalists to predict where, on the phase diagram
of the uniform system, their trapped system will be.  The behavior of
the visibility confirmed that the uniform Mott lobe is entered when
the center of the density profiles is in the Mott phase.

We have focused here on $d=1$.  However, the basic qualitative point
we wish to emphasize is valid in any dimension: a careful
consideration of the confining potential in addition to the number of
particles is essential for a meaningful comparison of the critical
couplings obtained in experiments with those of the homogeneous
system.

Supported under ARO Award W911NF0710576 with funds from the DARPA OLE
Program.  G.G.B. supported in part by the CNRS (France) PICS 18796. We
acknowledge very useful conversations with M. Rigol, J.K. Freericks
and B.L. Polisar. We thank L. Pollet for commenting on equilibration
problems in an early version of Fig.1b.


\begin{thebibliography}{40}
\bibitem{higher-dim} Similar considerations hold in higher dimensions,
with appropriate alterations. In $2d$ one has,
\begin{eqnarray}
N_b &=& 2\pi \int_0^{\infty} r dr
\, \rho_{2d}^0(\mu - V_T r^2 ; U);
\nonumber
\\
N_b {V_T} &\equiv& \tilde{\rho} = \pi \int_{-\infty}^{\mu} {d\mu_r
\, \rho_{2d}^0( \mu_r; U)}  \equiv I_2 (\mu; U).
\nonumber
\end{eqnarray}
Similarly, in $3d$,
\begin{eqnarray}
N_b &=& 4\pi \int_0^{\infty} r^2 dr
\rho_{3d}^0(\mu - V_T r^2 ; U);
\nonumber
\\
N_b {V_T}^{3/2} &\equiv& \tilde{\rho} = 2 \pi \int_{-\infty}^{\mu}
{d\mu_r \rho_{3d}^0( \mu_r; U)} {\sqrt{\mu-\mu_r}} \equiv I_3 (\mu;
U). \nonumber
\end{eqnarray}
\bibitem{rhomu}
Although $\rho(\mu)$ is monotonic, it of
course exhibits interesting features like a Mott plateau
where $\rho$ is constant for a range of $\mu$,
for $U$ sufficiently large, and a diverging derivative
$\kappa = \frac{\partial \rho}{\partial \mu}$ as the
Mott plateaus are approached \cite{fisher89,batrouni90}.
\end{thebibliography}
\end{document}